\documentclass[a4paper]{jpconf}
\usepackage{graphicx}
\usepackage{epsf}
\usepackage{amsfonts}
\usepackage{amsmath}
\usepackage{bm}
\usepackage{color}
\usepackage{slashed}
\newcommand{\be}{\begin{equation}}\newcommand{\ee}{\end{equation}}
\newcommand{\bea}{\begin{eqnarray}}\newcommand{\eea}{\end{eqnarray}}
\newcommand{\brr}{\begin{array}}\newcommand{\err}{\end{array}}
\newcommand{\bit}{\begin{itemize}}\newcommand{\eit}{\end{itemize}}
\newcommand{\ben}{\begin{enumerate}}\newcommand{\een}{\end{enumerate}}

\newcommand{\bbm}{\begin{bmatrix}}\newcommand{\ebm}{\end{bmatrix}}

\newcommand{\ba}{\begin{array}}
\newcommand{\ea}{\end{array}}
\newcommand{\G}{\textbf}

\newtheorem{mydef}{Definition}
\newtheorem{Lemma}{Lemma}
\newtheorem{theorem}{Theorem}
\newcommand{\bd}{\begin{mydef}} \newcommand{\ed}{\end{mydef}}
\newcommand{\bthe}{\begin{theorem}} \newcommand{\ethe}{\end{theorem}}
\newcommand{\ble}{\begin{Lemma}} \newcommand{\ele}{\end{Lemma}}

\newcommand{\dr}{\mathrm{d}}

\definecolor{darkred}{rgb}{.8,0,0}

\definecolor{darkblue}{rgb}{0,0,.7}

\def\intx{\int \!\!\mathrm{d}^3 x}
\def\intk{\int \!\!\mathrm{d}^3 k}

\def\ph{\varphi}
\def\lan{\langle}
\def\lf{\left}

\def\non{\nonumber}\def\pa{\partial}\def\ran{\rangle}

\def\ri{\right}
\def\al{\alpha}\def\bt{\beta}\def\ga{\gamma}
\def\de{\delta}\def\De{\Delta}

\def\la{\lambda}\def\si{\sigma}
\def\om{\omega}\def\Om{\Omega}

\def\1{{_{1}}}\def\2{{_{2}}}

\newcommand{\ide}{1\hspace{-1mm}{\rm I}}

\def\noHe0{:\;\!\!\;\!\!:H_e(0):\;\!\!\;\!\!:}
\def\noHm0{:\;\!\!\;\!\!:H_\mu(0):\;\!\!\;\!\!:}

\def\lan{\langle}
\def\lf{\left}

\def\non{\nonumber}
\def\pa{\partial}\def\ran{\rangle}

\def\ri{\right}

\def\al{\alpha}\def\bt{\beta}\def\ga{\gamma}
\def\de{\delta}\def\De{\Delta}

\def\la{\lambda}
\def\si{\sigma}
\def\om{\omega}\def\Om{\Omega}

\def\1{{_{1}}}\def\2{{_{2}}}

\begin{document}
\title{Flavor neutrino states for pedestrians}

\author{M Blasone$^{1}$, L Smaldone$^{1}$ and G Vitiello$^{1}$}
\address{$^1$ Dipartimento di Fisica, Universit\`a di Salerno, Via Giovanni Paolo II, 132 84084 Fisciano, Italy \& INFN Sezione di Napoli, Gruppo collegato di Salerno, Italy}

\ead{blasone@sa.infn.it}
\ead{lsmaldone@sa.infn.it}
\ead{vitiello@sa.infn.it}

\begin{abstract}
In this paper we discuss the ontology of flavor states of oscillating neutrinos. While an heuristic approach to this subject, experimentally successful in the high energy regime, is generally adopted, a logically consistent definition of flavor states describing neutrinos produced and detected in weak processes is still desirable and essential from a theoretical perspective. Here we briefly review basic facts and  present some arguments which suggest that the definition of flavor states as eigenstates of flavor charges is the most reasonable one. 
\end{abstract}

\section{Introduction}
Neutrinos are strange beasts which changed  identity several times during their history: they were firstly theorized by Pauli to explain the continuous spectrum of electrons in the $\beta$ decay\footnote{The name \emph{neutrino} was introduced by E.~Fermi, to distinguish it from the neutron \cite{Seg}.}. They were observed for the first time, after long efforts, by F. Reines and C. Cowan \cite{ReiCow}. Then it was proved that neutrinos exist in three different \emph{flavors} \cite{mutau}, in correspondence with the three flavors of charged leptons.

For almost three decades people thought that neutrinos were exactly massless, and then they were described as left-handed Weyl fermions \cite{Weyl}. This assumption is also contained in the original Standard Model (SM) by Weinberg and Salam \cite{SM}. However, the results of the Homestake experiment~\cite{Davis}, and many others along the years~\cite{neutrinoexp}, inescapably proved that they must possess small masses. Actually, the nature itself of neutrino masses, Dirac \cite{Dirac} or Majorana \cite{Majorana} is not clear at the moment and different extensions of SM \cite{seesaw,ChengLi1980,King2008} were proposed. 

Neutrinos also possess flavor mixing, a feature which is at the basis of the phenomenon of neutrino  oscillations. This was theorized by Pontecorvo~\cite{Pontecorvo}, Maki, Nakagawa and Sakata \cite{sakata}, and it is nowadays universally accepted~\cite{BilPet,partref}. 

Although quantum mechanical (QM) Pontecorvo theory works extremely well in the actual experiments where neutrinos, due to their small masses, are ultrarelativistic, it is not satisfactory from a theoretical point of view, and deviations in low energy experiments are expected in different scenarios based on quantum field theory (QFT) \cite{Man,GKL92,Blasone:1995zc,NeutPheno,qftmixing}. In particular, the correct definition of flavor states is somewhat mysterious. The reason is that, because of the flavor oscillation phenomenon, these cannot be constructed as the usual $in$ or $out$ states \cite{LSZ}. A good review of the different approaches can be found in Ref. \cite{MixingReview}.

In the present paper we review few basic facts on neutrino flavor states. In doing so we will limit to the case of Dirac neutrinos. We also show some paradoxes and mistakes that may occur in an oversimplified treatment of neutrino mixing and oscillations and propose some arguments which suggests that a correct description of neutrino flavor states would be the one developed in Refs. \cite{Blasone:1995zc,qftmixing}.

\section{Pontecorvo flavor states} \label{bfmixing}
The mixing transformation is 
\be
\nu_\si(x) \ = \ \sum_{j} \, U_{\si \, j} \nu_j(x) \, , 
\ee
where $\nu_\si$ are the \emph{flavor fields}, involved in the weak interaction, $\nu_j$ are the mass fields, which describe neutrinos with definite masses and $U$ is the unitary mixing matrix. 

The problem of constructing a Fock space for flavor neutrinos was studied many times in literature~\cite{Man,GKL92,Blasone:1995zc,qftmixing,Ho2012,Lobanov}. There is general agreement~\cite{GKL92,Blasone:1995zc,qftmixing,Ho2012,Lobanov,Fantini:2018itu} on the existence of such a Fock space in the ultra-relativistic limit $m_j/|\G k| \rightarrow 0$, where $m_j$ are the neutrino masses. In this case annihilation operators are defined as
\be
\tilde{\al}^r_{\G k, \si} \ = \ \sum_{j} \, U^*_{\si \, j} \, \al^r_{\G k, j}
\ee
where $\al^r_{\G k, j}$ are the annihilation operator of fields with definite mass:
\begin{eqnarray}
\nu _{j}(x) =  \sum_r \,  \int \!\! \frac{\dr^3 k}{(2 \pi)^{\frac{3}{2}}} \, \left[ u_{{\bf k},j}^{r}(t) \, \alpha _{{\bf k},j}^{r} +  \ v_{-{\bf k},j}^{r}(t)  \,\beta _{-{\bf k},j}^{r\dagger
}\right]  e^{i{\bf k}\cdot {\bf x}}  \, .
\label{fieldex}
\end{eqnarray}
Similar relations hold for $\beta _{{\bf k},j}^{r}$. Flavor states can be thus constructed as:
\be \label{mv}
|\nu^r_{\G k,\si} \ran_{_P} \ \equiv \ \tilde{\al}^{r\dag}_{\G k, \si}|0\ran \, ,
\ee
where $|0\ran$ is the vacuum state, which is annihilated by $\al^{r}_{\G k, j}$ and $\bt^{r}_{\G k, j}$ (\emph{mass vacuum}). These states can be recognized as the flavor states originally introduced by Pontecorvo~\cite{Pontecorvo}:
\be
|\nu^r_{\G k,\si} \ran_{_P} \ = \  \sum_{j} \, U^*_{\si \, j} \, |\nu^r_{\G k,j} \ran \, .
\ee
In the relativistic limit these are eigenstates of flavor charges at fixed time:
\be \label{relei}
\lim_{m_i/|\G k|\rightarrow 0} \, Q_{\nu_\si}(0)|\nu^r_{\G k,\si} \ran_{_P} \ = \ |\nu^r_{\G k,\si} \ran_{_P} \, ,
\ee
where \cite{ChargesBJV,Nishi2006} (see Section \ref{lepnum})
\be
Q_{\nu_\si}(t)  \ \equiv \ \intx \, \nu^\dag_\si(x) \, \nu_\si(x) \, .
\ee
However, this is not true at all energy scales. To see this, let us explicitly consider the 2-flavor case ($\si=e,\mu$ and $j=1,2$). The mixing matrix is given by:
\be
U \ = \ \begin{pmatrix} \cos \theta & \sin \theta \\ -\sin \theta & \cos \theta \end{pmatrix} \, .
\ee
We can evaluate the oscillation formula as the expectation value of the flavor charge on a reference neutrino state \cite{ChargesBJV,BHV99,Nishi2006}:
\be
\widetilde{\mathcal{P}}_{e\rightarrow \mu}(t) \ \equiv \ {}_{_P}\lan \nu^r_{\G k,e}|Q_{\nu_\mu}(t)|\nu^r_{\G k,e} \ran_{_P} \ = \ \frac{\sin^2 (2 \theta)}{2}\Big\{1-|U_\G k| \, \cos [(\om_{\G k,1}-\om_{\G k,2})t]\Big\} \, ,
\ee
where $|U_\G k| \equiv u^{r\dag}_{\G k,1}u^{r}_{\G k,2}$. In the ultra-relativistic case $|U_\G k| \rightarrow 1$, and we get the standard oscillation formula:
\be 	\label{stafor}
\mathcal{P}_{e\rightarrow \mu}(t) \ = \ \sin^2 (2 \theta)\,\sin^2 \lf(\frac{\om_{\G k,1}-\om_{\G k,2}}{2}t\ri) \, .
\ee
Note in particular that 
\be
\widetilde{\mathcal{P}}_{e\rightarrow \mu}(0) \ = \ \frac{\sin^2 (2 \theta)}{2}\lf(1-|U_\G k|\ri) \, ,
\ee
which is unacceptable because it tells us that flavor is undefined even at $t=0$~\cite{Nishi2006,Chargeei}. A similar paradox was also pointed out in Ref.~\cite{BHV99}, in connection with the study of Feynman propagator for flavor fields. 

\section{Lepton number conservation} \label{lepnum}

\subsection{Basic considerations}

In Ref.~\cite{GKL92} it was pointed out that the amplitude of the neutrino detection process $\nu_\si+X_i \rightarrow e^- + X_f$, where $X_i$ and $X_f$ are the initial and the final particles, respectively, and $e^-$ is the electron, is generally different from zero if $\si \neq e$, if we use the Pontecorvo states. In fact, for low-energy weak processes (where we can use the four-fermion Fermi interaction):
\be \label{wae}
\lan e^{s}_{\G q,-}|\bar{e}(x) \, \ga^\mu \, (1-\ga^5) \, \nu_e(x)|\nu^r_{\G k,\si} \ran_{{}_P} \, h_\mu(x) \ = \ \sum_j \, U_{e j} 	\, U^*_{\si j} \lan e^{s}_{\G q,-}|\bar{e}(x) \, \ga^\mu \, (1-\ga^5) \, \nu_j(x)|\nu^r_{\G k,j} \ran \, h_\mu(x) \, , 
\ee
where $h_\mu$ are the matrix elements of the $X$ part. This is generally different from $\de_{\si e}$. This seems to be inconsistent because the flavor of the neutrino is \emph{defined} by the flavor of the associated charged lepton in the (lepton-neutrino) doublet (see also Ref. \cite{BilGiu}). From this observation, in the attempt of developing a consistent QFT approach, \emph{weak process states} were introduced\footnote{Here we are presenting the so called \emph{detection states}. Analogously \emph{production} states, can be defined~\cite{GKL92}. } \cite{GKL92}:
\be 	\label{wp}
|\nu^r_\si \ran_{{}_{WP}} \ = \ \frac{1}{\sqrt{\sum_{j}|\mathcal{A}_{\si j}|^2}} \, \sum_{j} \, \mathcal{A}_{\si j} \, |	\nu^r_{{\G k}_j,j} 	\ran \, ,  
\ee
where
\be
\mathcal{A}_{\si j} \ = \ \lan \nu^r_{{\G k}_j,j} X^{r'}_{\G k',i} |S^\dag | l^{s,-}_{\G q,\si} \, X^{s'}_{\G q',f} \ran \, . 
\ee
Here $l^-_\al$ indicates the charged lepton and $S$ is the $S$-matrix. If we consider, as before, a low-energy weak process, we can write:
\be
\mathcal{A}_{\si j} \ = \ U^*_{\si j} \, \mathcal{M}_{\si j} \, , 
\ee
with
\be
\mathcal{M}_{\si j} \ \equiv \ -i \frac{G_F}{\sqrt{2}} \, \int \!\! \dr^4 x \, \lan \nu^r_{{\G k}_j,j} X^{r'}_{\G k',i}|\bar{\nu}_j(x) \, \ga^\mu \, (1-\ga^5) \, l_\si(x) J_\mu(x)|l^{s,-}_{\G q,\si} X^{s'}_{f,\G q'} \ran  \, ,
\ee
where $J_\mu(x)$ is the $X$ current. It is important to stress that the definition~\eqref{wp} explicitly depends on the details of the detection process. However, the inconsistency of Eq.\eqref{wae} was not solved by this new definition. In fact:  
\be
\lan e^{s}_{\G q,-}|\bar{e}(x) \, \ga^\mu \, (1-\ga^5) \, \nu_e(x)|\nu^r_\si \ran_{{}_{WP}} \, h_\mu(x)  =  \sum_j  \frac{U_{e j} 	\, \mathcal{A}_{\si j}}{\sqrt{\sum_{j}|\mathcal{A}_{\si j}|^2}} \lan e^{s}_{\G q,-}|\bar{e}(x) \, \ga^\mu \, (1-\ga^5) \, \nu_j(x)|\nu^r_{{\G k}_j,j} \ran \, h_\mu(x) \, , 
\ee
which is also generally different from $\de_{\si e}$. 

We now have to do an important remark. The $S$ matrix is defined so it connects $in$ and $out$ states \cite{LSZ}:
\be
S_{AB} \ \equiv \ \lan  A; out |B; in\ran \, .
\ee
The asymptotic states are defined far before and after the interaction, so that neutrino oscillation occurs leading to the violation of family lepton number conservation. This is natural and expected. The previous example are thus not pathological in this sense. However, lepton number has to be conserved in the production and detection vertexes (at tree level), where flavor oscillation can be neglected\footnote{Obviously loop diagrams can produce violation of lepton number in the production and detection vertexes, but these contributions are negligible in the following discussion.}. Because this point created some discussion in literature~\cite{LiLiu,Blasone:2006jx,Lee}, we think it is important to expose it in more detail, starting from basic elementary considerations in order to avoid to incur in the contradictions of oversimplified treatments. 

\subsection{Flavor charge conservation in the vertex}
Let us consider a charged weak decay  $W^+\rightarrow e^+ + \nu_e$. The relevant part of SM Lagrangian is 
$\label{Lagrangian} \mathcal{L}=\mathcal{L}_0+\mathcal{L}_{int}$ with
\bea
&&{\cal L}_{0}(x)  =  \overline{\nu}(x)  \lf( i \ga_\mu \pa^\mu - M_{\nu} \ri)\nu(x) \, + \, \overline{l}(x) \lf( i \ga_\mu \pa^\mu - M_{l} \ri) l(x) \, , \\[2mm]
 &&{\cal L}_{int}(x)  =  \frac{g}{2\sqrt{2}}
\lf [ W_{\mu}^{+}\,
\overline{\nu}(x)\,\gamma^{\mu}\,(1-\gamma^{5})\,l(x) +
h.c. \ri] \, ,
\label{L-interact}
\eea
where $\nu = \lf(\nu_e, \nu_\mu \ri)^{{T}}  , \, l = \lf(e, \mu \ri)^T$, and
\bea \label{neutmass}
&&\mbox{\hspace{-5mm}} M_{\nu}\,=\,  \lf(\ba{cc}m_e & m_{e\mu}
\\ m_{e\mu} & m_\mu\ea\ri), \;\;\;\;\;
M_l\,=\,  \lf(\ba{cc}\tilde{m}_e &0 \\
0 & \tilde{m}_\mu\ea\ri) .
\eea
The Lagrangian $\mathcal{L}$ is invariant under the global $U(1)$ transformations
$\nu \rightarrow e^{i \alpha} \nu$ and $l \rightarrow e^{i \alpha} l$
leading to the conservation of the total  flavor charge $Q_{l}^{tot}$ corresponding to the total lepton-number conservation~\cite{BilPet,BJSun}. This can be written in terms of the flavor charges for neutrinos and charged leptons~\cite{ChargesBJV}
\be
Q_{l}^{tot} =  \sum_{\si=e,\mu} Q_\si^{tot}(t) \,,\quad   Q_{\si}^{tot} (t) = Q_{\nu_{\si}}(t) + Q_{\si}\,,
\ee
with
\bea
Q_{e} & = &  \intx \,
e^{\dag}(x)e(x) \,, \qquad Q_{\nu_{e}} (t) =  \intx \,
\nu_{e}^{\dag}(x)\nu_{e}(x)\,,
\nonumber \\ [1mm]
Q_{\mu} &  = &   \intx \,
 \mu^{\dag}(x) \mu(x)\,, \qquad Q_{\nu_{\mu}} (t)= \intx \, \nu_{\mu}^{\dag}(x) \nu_{\mu}(x)\,  .
 \label{QflavLept}
\eea
The above charges can be derived via Noether's theorem~\cite{Noether}  from the Lagrangian~\eqref{L-interact}. Note the time dependence of the neutrino charges, due to the non-diagonal mass matrix $M_{\nu}$.

 By observing that  $[\mathcal{L}_{int}({\bf x},t),Q_\si^{tot}(t)]=0$, we see that a neutrino flavor state is well defined in the production vertex
as an eigenstate of the corresponding  flavor charge~\cite{Chargeei}. Let us now consider the (Pontecorvo states) amplitude of the above mentioned process
\be
\mathcal{A}^P_{W^+ \rightarrow e^+ \, \nu_e} \ = \ {}_{_P}\lan \nu^r_{\G k, e}| \otimes \lan e^s_\G q |\lf[-i \int^{x^0_{out}}_{x^0_{in}} \, \dr^4 x \, \mathcal{H}^e_{int}(x) \ri]|W^+_{\G p, \la} \ran \, .
\ee
The interaction Hamiltonian density is
\be
\mathcal{H}^e_{int}(x) \ = \  -\frac{g}{2\sqrt{2}} \, W^+_\mu(x) \, J^\mu_e(x) + h.c. \, , 
\ee
and  
\be
J^\mu_e(x) \ = \ \bar{\nu}_e(x) \, \ga^\mu \, (1-\ga^5)e(x) \, , 
\ee
as it can be deduced from Eq.\eqref{L-interact}. The usual amplitude is obtained by taking the asymptotic limit $x_{out}^0 \rightarrow + \infty$, $x_{in}^0 \rightarrow - \infty$. However, as mentioned, the flavor states are not asymptotic stable states, and we want to investigate the short-time behavior of the amplitude (around the interaction time $x_0=0$). Explicit calculations (cfr. Ref.\cite{Blasone:2006jx}) give:
\bea
\mathcal{A}^P_{W^+ \rightarrow e^+ \, \nu_e} & = & \frac{i g }{2 \sqrt{4 \pi}}\frac{\varepsilon_{\G p, \mu, \la}}{\sqrt{2 E^W_p}}\, \de^3(\G p-\G q -\G k) \non \\[2mm]
& \times & \sum^2_{j=1} \, U^2_{e j} \, \int^{x^0_{out}}_{x^0_{in}} \!\! \dr x^0 \, e^{-i \om_{\G k,j} \, x^0_{out}} \, \bar{u}^r_{\G k,j} \, \ga^\mu (1-\ga^5) \, v^s_{\G q,e} \, e^{-i (E^W_p-E_q^e-\om_{\G k,j}) x^0} \, , 
\eea
where $\varepsilon_{\G p, \mu, \la}$ is the polarization vector of $W^+$ and $v^s_{\G q,e}$ is the positron wave function. We take $x^0_{in}=-\De t /2$ and $x^0_{out}=\De t /2$, when $\tau_W \ll \De t \ll t_{osc}$, where $\tau_W$ is the $W^+$ lifetime, while $t_{osc}$ is the oscillation time. Under this condition, we can expand the amplitude at the leading order in $\De t$, obtaining:
\bea
\mathcal{A}^P_{W^+ \rightarrow e^+ \, \nu_e} & \approx & \frac{i g }{2 \sqrt{4 \pi}}\frac{\varepsilon_{\G p, \mu, \la}}{\sqrt{2 E^W_p}}\, \de^3(\G p-\G q -\G k) \De t \, \sum^2_{j=1} \, U^2_{e j} \ \, \bar{u}^r_{\G k,j} \, \ga^\mu (1-\ga^5) \, v^s_{\G q,e}  \, . 
\eea
In the same way, one can evaluate the ``wrong'' amplitude
\be
\mathcal{A}^P_{W^+ \rightarrow e^+ \, \nu_\mu} \ = \ {}_{_P}\lan \nu^r_{\G k, \mu}| \otimes \lan e^s_\G q |\lf[-i \int^{x^0_{out}}_{x^0_{in}} \, \dr^4 x \, \mathcal{H}^e_{int}(x) \ri]|W^+_{\G p, \la} \ran \, .
\ee
Proceeding as before we get:
\bea
\mathcal{A}^P_{W^+ \rightarrow e^+ \, \nu_\mu} & = & \frac{i g }{2 \sqrt{4 \pi}}\frac{\varepsilon_{\G p, \mu, \la}}{\sqrt{2 E^W_p}}\, \de^3(\G p-\G q -\G k) \non \\[2mm]
& \times & \sum^2_{j=1} \, U_{\mu j} \, U_{e j} \, \int^{x^0_{out}}_{x^0_{in}} \!\! \dr x^0 \, e^{-i \om_{\G k,j} \, x^0_{out}} \, \bar{u}^r_{\G k,j} \, \ga^\mu (1-\ga^5) \, v^s_{\G q,e} \, e^{-i (E^W_p-E_q^e-\om_{\G k,j}) x^0} \, , 
\eea
which in the short time limit becomes
\bea
\mathcal{A}^P_{W^+ \rightarrow e^+ \, \nu_\mu}  \approx  \frac{i g }{2 \sqrt{4 \pi}}\frac{\varepsilon_{\G p, \mu, \la}}{\sqrt{2 E^W_p}}\, \de^3(\G p-\G q -\G k) \De t \, \sum^2_{j=1} \, U_{\mu j} \, U_{e j} \ \, \bar{u}^r_{\G k,j} \, \ga^\mu (1-\ga^5) \, v^s_{\G q,e}  \, . 
\eea
This is clearly different from zero, which is inconsistent. 

Let us notice that the same procedure for the weak process states above defined would lead to
\bea
\mathcal{A}^{{}_{WP}}_{W^+ \rightarrow e^+ \, \nu_\mu}  \approx  \frac{i g }{2 \sqrt{4 \pi}}\frac{\varepsilon_{\G p, \mu, \la}}{\sqrt{2 E^W_p}}\, \De t \, \sum^2_{j=1} \, \frac{\de^3(\G p-\G q -\G k_j)\, \mathcal{A}^*_{\mu j}}{\sqrt{\sum^2_{j=1}|\mathcal{A}_{\mu j}|^2}} \, U_{e j} \ \, \bar{u}^r_{{\G k}_j,j} \, \ga^\mu (1-\ga^5) \, v^s_{\G q,e}  \, . 
\eea
The main reason behind this ambiguity is that Eq.\eqref{relei} does not hold in the non-relativistic case. The same is true for weak process states: these are \emph{not} eigenstates of the flavor charges which commute with the interaction vertex. Therefore it is obvious that the use of these states leads to violation of the family lepton number at tree level, in the production (and detection) vertex. 
\section{First quantized oscillation formula and Dirac equation}
In this section, following Ref.~\cite{Alex2005} we review the correct derivation of the oscillation formula in relativistic quantum mechanics. Flavor wavefunction satisfies the Dirac equation:
\be
\lf(i \ga^\mu \, \pa_\mu \otimes \ide_2- \ide_4 \otimes M_\nu \ri) \, \Psi(x) \ = \ 0 \, , 
\ee
where $\ide_n$ indicates the $n \times n$ identity matrix and $M_\nu$ is the mass matrix introduced in Eq.\eqref{neutmass}.

For simplicity we limit to study the problem in one spatial dimension (along the $z$-axis). By introducing wavefunctions of neutrino with definite masses:
\be
\lf(i \ga^0 \pa_0+i \ga^3 \pa_3-m_j\ri) \, \psi_j(z,t) \ = \ 0 \,, \qquad j=1,2 \, , 
\ee
one can express a neutrino wavepacket $\Psi$ as:
\bea
\Psi(z,t) & = & \cos \theta \, \psi_1(z,t) \, \otimes \,\nu_1 \, + \, \sin \theta \, \psi_2(z,t) \, \otimes \, \nu_2 \non \\[2mm]
& = & \lf[\psi_1(z,t) \, \cos^2 \theta \, + \, \psi_2(z,t) \, \sin^2 \theta \ri] \, \otimes \, \nu_\si + \sin \theta \, \cos \theta \, \lf[\psi_1(z,t)-\psi_2(z,t)\ri]\, \otimes \, \nu_\rho \non \\[2mm] \label{flavfunc}
& \equiv & \psi_\si(z,t) 	\, \otimes \, \nu_\si \, + \, \psi_\rho(z,t) \, \otimes \, \nu_\rho \, , 
\eea
where $\psi_j(z,t)$ are the wavepackets describing neutrinos with definite masses, $\nu_1,\nu_2$ are the eigenstates of $M_\nu$ and $\nu_\si, \nu_\rho$ are flavor eigenstates. A neutrino is produced as a flavor eigenstate if $\psi_1(z,0)= \psi_2(z,0)=\psi_\si(z,0)$, with $\si=e,\mu$. The oscillation probability will be given by
\be
P_{\nu_\si \rightarrow \nu_\rho} \ = \ \int^{+\infty}_{-\infty} \!\! \dr z \, \psi^\dag_\rho(z,t) \, \psi_\rho(z,t) \, .
\ee
By using Eq.\eqref{flavfunc} we can derive:
\be
P_{\nu_\si \rightarrow \nu_\rho} \ = \ \frac{\sin^2 2 \theta}{2} \, \lf[1-I_{{}_{12}}(t)\ri] \, , 
\ee
where the interference term is given by
\be 	\label{interf}
I_{{}_{12}}(t) \ = \ \Re e \lf[\int^{+\infty}_{-\infty} \!\! \dr z \, \psi^\dag_1(z,t) \, \psi_2(z,t)\ri] \, .
\ee
Note that respect to the usual treatment, where only positive frequency modes are included, this analysis explicitly shows that also negative frequency contribution have to be involved in the computation of the interference term~\eqref{interf}. In fact, let us consider the Fourier expansion of $\psi_j(z,t)$:
\begin{eqnarray}
\psi _{j}(x) =  \sum_r \, \int^{+\infty}_{-\infty} \!\! \frac{\dr p_z}{2 \pi} \, \left[ u_{p_z,j}^{r} \, \alpha _{p_z,j}^{r} \, e^{-i \, \om_{p_z,j} \, t} +  \ v_{-p_z,j}^{r}  \beta _{-p_z,j}^{r*
}\, e^{i \, \om_{p_z,j} \, t}\right]  e^{i \, p_z \, z} \, ,  \quad j=1,2 \, .
\label{psiex}
\end{eqnarray}
The requirement that neutrino is produced with definite flavor, assumes the form:  
\be
u_{p_z,j}^{r} \, \alpha_{p_z,j}^{r}  +  \ v_{-p_z,j}^{r}  \beta_{-p_z,j}^{r*
} \ = \ \ph_\si(p_z-p_0) \, w \, , 
\ee
where $\ph_\si(p_z-p_0)$ is the flavor neutrino distribution in the momentum space, at $t=0$, $p_0$ is the mean momentum of mass wavepackets and $w$ is a constant spinor, satisfying $w^\dag w =1$. By using orthogonality conditions of Dirac spinors we derive the relations
\bea
\alpha_{p_z,j}^{r} & = & \ph_\si(p_z-p_0) \, u_{p_z,j}^{r\dag} \, w \, , \\[2mm]
\bt_{-p_z,j}^{r*} & = & \ph_\si(p_z-p_0) \, v_{-p_z,j}^{r\dag} \, w \, .
\eea
Substituting in Eq.\eqref{psiex} and then in Eq.\eqref{interf} we finally arrive at~\cite{Nishi2006,Alex2005}:
\be
I_{{}_{12}}(t) \ = \ \int^{+\infty}_{-\infty} \!\! \frac{\dr p_z}{2 \pi} \, \ph^2_\si(p_z-p_0) \, \lf(|U_{p_z}|^2 \, \cos (\Om^-_{p_z} t) \, + \, |V_{p_z}|^2 \, \cos (\Om^+_{p_z} t)\ri) \, , 
\ee
where
\bea
\Om^{_\pm}_{p_z} & = & \om_{p_z,1} \pm \om_{p_z,2} \, , \\[2mm]
|V_{p_z}|^2 & = & 1-|U_{p_z}|^2 \ = \ \frac{\om_{p_z,1} \, \om_{p_z,2}-p_z^2-m_1 m_2}{2 \om_{p_z,1} \, \om_{p_z,2}} \, . 
\eea
The notation here is slightly different respect to Refs.\cite{Nishi2006,Alex2005}, in order to get in touch with the next section. For plane waves, $\ph_\si(p_z-p_0) \ = \ \de(p_z-p_0)$. The oscillation probability thus reads
\be \label{fqosc}
P_{\nu_\si \rightarrow \nu_\rho} \ = \ \sin^2 2 \theta \, \lf[|U_{p_0}|^2 \, \sin^2 \lf(\frac{\Om^{_-}_{p_z}}{2} t\ri) \, + \, |V_{p_0}|^2 \, \sin^2 \lf(\frac{\Om^{_+}_{p_0}}{2} t\ri)\ri] \, .
\ee
The main difference with respect to the standard oscillation formula Eq.\eqref{stafor} is the presence of a fast oscillating term, with frequency $\Om^{_+}_{p_0}/2$. This is analogous to the \emph{Zitterbewegugng} encountered in atomic physics, which leads to the Darwin contribution to fine structure of hydrogen atom. In our case this effect is very small when $p_0 \gg \sqrt{m_1 m_2}$, i.e. for ultrarelativistic neutrinos. In that regime $|U_{p_0}|^2 \rightarrow 1$ and $|V_{p_0}|^2 \rightarrow 0$, and the oscillation probability recovers its standard form~\eqref{stafor}. 

Let us remark that the usual treatment~\cite{Pontecorvo,BilPet,partref}, not including negative frequency terms, is very similar to the \emph{rotating wave approximation} usually encountered in quantum optics and atomic physics \cite{Kurcz:2010wx,Fuji2017}, where fast oscillating terms in the Hamiltonian are neglected in order to find exact solutions of the eigenvalue problem. However,  in the case of neutrino oscillations, there are no reasons to neglect the contribution with $\Om^+_{p_0}$, apart from ultrarelativistic case. 
\section{Flavor eigenstates}
Let us now come back to the problem of constructing flavor states in QFT. Resuming to our previous considerations, we aim to construct states which are eigenstates of the flavor charges. To this end, we limit ourselves to the two flavor case. The mixing relations for neutrino fields are
\bea
\left(
  \begin{array}{c}
    \nu_e (x) \\
    \nu_{\mu}(x) \\
  \end{array}
\right) \ = \ \left(
                \begin{array}{cc}
                  \cos\theta & \sin\theta \\
                  -\sin\theta & \cos\theta \\
                \end{array}
              \right)\left(
                       \begin{array}{c}
                        \nu_1 (x)\\
                         \nu_2 (x) \\
                       \end{array}
                     \right), \label{PontecorvoMix1}
\eea
with $\tan 2 \theta = 2 m_{e\mu}/(m_\mu-m_e)$. Eq. (\ref{PontecorvoMix1}) can be equivalently rewritten as~\cite{qftmixing}
\bea
\label{MixingRel1}
\nu_\si(x)=
G_{\theta}^{-1}(t)\, \nu_j (x) \,  G_{\theta}(t) \, , \label{MixingRel2}
\eea
with $(\si,j)=(e,1),(\mu,2)$ and  $G_\theta(t)$  given by
\be
\mbox{\hspace{-2mm}}G_{\theta}(t)=\exp\left[\theta\intx \, \left(\nu_1^{\dagger}(x)\nu_2(x)-
\nu_2^{\dagger}(x)\nu_1(x)\right)\right].
\label{MixGen}
\ee
From (\ref{fieldex}) and (\ref{MixingRel1}) it follows that flavor fields are:
\begin{eqnarray}
\nu _{\si}(x) \ = \ \sum_r \,  \int \!\! \frac{\dr^3 k}{(2 \pi)^{\frac{3}{2}}}\!\! \ \,
\left[ u_{{\bf k},j}^{r}(t) \, \alpha _{{\bf
k},\si}^{r}(t) \, + \, v_{-{\bf k},j}^{r}(t) \, \beta _{-{\bf k},\si}^{r\dagger
}(t)\right] \, e^{i{\bf k}\cdot {\bf x}} \, ,
\label{fieldex1}
\end{eqnarray}
with $(\si,j) \ = \ (e,1), (\mu,2)$, and flavor ladder operators are given by
\bea\label{flava}
\begin{pmatrix}
\alpha^{r}_{{\bf
k},\si}(t)\\
\beta^{r}_{-{\bf k},\si}(t)
\end{pmatrix}
\ = \
G_{\theta}^{-1}(t)\,
\begin{pmatrix}
\alpha^{r}_{{\bf
k},j}(t)\\
\beta^{r}_{-{\bf k},j}(t)
\end{pmatrix}
\,  G_{\theta}(t) \, .
\eea
The \emph{flavor vacuum} is defined as \cite{Blasone:1995zc,qftmixing}:
\be
\label{flavvac} |0\rangle_{e,\mu} = G^{-1}_{\theta}(0)\;
|0 \rangle_{1,2}\; , 
\ee
where $|0\ran_{1,2}$ denotes the mass vacuum (cf. Eq.\eqref{mv}) for the two-flavor case. 																			
One can easily verify that it is annihilated by the flavor operators
defined in Eq.~\eqref{flava}. Moreover, one can prove that \cite{Blasone:1995zc,qftmixing}
\be \label{ineqrep}
\lim_{V \rightarrow \infty}{}_{1,2} \lan 0|0 \rangle_{e,\mu} =  \lim_{V \rightarrow \infty} e^{\frac{V}{(2\pi)^3}\intk \ln(1-\sin^2 \theta|V_\G k|^2)^2}  =  0 \, ,
\ee
where
\bea
&&\mbox{\hspace{-9mm}}|V_\G k|  = \frac{|\G k|}{\sqrt{4 \om_{\G k,1}\om_{\G k,1}}}
\lf(\sqrt{\frac{\om_{\G k,2}+m_2}{\om_{\G k,1}+m_1}}-\sqrt{\frac{\om_{\G k,1}+m_1}{\om_{\G k,2}+m_2}}\ri)\!\!,
\eea
i.e. flavor and massive fields belong to unitarily inequivalent representations of the anticommutation relations~\cite{Umezawa,BJV}. 

True flavor eigenstates can be explicitly constructed as
\be \label{bvflavstate}
|\nu^r_{\G k,\si}\ran \ = \ \al^{r\dag}_{\G k,\si} |0\ran_{e ,\mu}  \, .
\ee
where flavor operators are taken at reference time $t=0$. One can prove that:
\be
Q_{\nu_\si}(0) |\nu^r_{\G k,\si}\ran \ = \ |\nu^r_{\G k,\si}\ran \, .
\ee
The corresponding oscillation formula can be found by taking the expectation value of the flavor charges~\cite{BHV99}
\be
\mathcal{Q}_{\si\rightarrow \rho}(t) \ = \ \lan Q_{\nu_\rho}(t) \ran_\si \, ,
\ee
where $\langle \cdots\rangle_\si = \lan \nu^r_{\G k,\si}| \cdots |\nu^r_{\G k,\si}\ran$, which gives
\bea \label{oscfor}
&& \mbox{\hspace{-4mm}}\mathcal{Q}_{\si\rightarrow \rho}(t)  =   \sin^2 (2 \theta)\Big[|U_\G k|^2\sin^2\lf(\frac{\Om_{\G k}^{_-}}{2}t\ri)+  |V_\G k|^2\sin^2\lf(\frac{\Om_{\G k}^{_+}}{2}t\ri)\Big]  , \nonumber \\[1mm]
&& \mbox{\hspace{-4mm}}\mathcal{Q}_{\si\rightarrow \si}(t)  =  1 \ - \ \mathcal{Q}_{\si\rightarrow \rho}(t) \, , \quad \si \neq \rho \, ,
\eea
where now $\Om_{\G k}^{_{\pm}}\equiv \om_{\G k,2}\pm\om_{\G k,1}$ and $|U_\G k|^2=1-|V_\G k|^2$. This is nothing but the formula~\eqref{fqosc}, obtained in the previous section from Dirac equation in the quantum mechanical wave packet formalism. 

By using the flavor states~\eqref{bvflavstate} we can also evaluate the amplitudes of the decay $W^+ \rightarrow e^+ \, \nu_e$ \cite{Blasone:2006jx}:
\bea \non 
\mathcal{A}_{W^+ \rightarrow e^+ \, \nu_e} & = & \frac{i g }{2 \sqrt{2}(2\pi)^\frac{3}{2}} \, \de^3(\G p-\G q -\G k) \, \int^{x^0_{out}}_{x^0_{in}} \!\! \dr x^0 \frac{\varepsilon_{\G p, \mu, \la}}{\sqrt{2 E^W_p}}\, \de^3(\G p-\G q -\G k) \\[2mm] \non
& \times &  \lf\{ \cos^2 \theta\, e^{-i \om_{\G k,1} \, x^0_{in}} \, \bar{u}^r_{\G k,1} \, \ga^\mu (1-\ga^5) \, v^s_{\G q,e} \, e^{-i (E^W_p-E_q^e-\om_{\G k,1}) x^0} \ri.	\\[2mm]
& + &\sin^2 \theta 	\lf[|U_\G k| \, e^{-i \om_{\G k,2} \, x^0_{in}} \, \bar{u}^r_{\G k,2} \, \ga^\mu (1-\ga^5) \, v^s_{\G q,e} \, e^{-i (E^W_p-E_q^e-\om_{\G k,2}) x^0}\ri. \non\\[2mm]
 & + & \lf.\lf. \varepsilon^r |V_\G k| \, e^{i \om_{\G k,2} \, x^0_{in}} \, \bar{v}^r_{-\G k,2} \, \ga^\mu (1-\ga^5) \, v^s_{\G q,e} \, e^{-i (E^W_p-E_q^e+\om_{\G k,2}) x^0}\ri]\ri\}\, ,
\eea
with $\varepsilon^r \equiv (-1)^r$, and the ``wrong'' one, $W^+ \rightarrow e^+ \, \nu_\mu$:
\bea \non 
\mathcal{A}_{W^+ \rightarrow e^+ \, \nu_\mu} & = & \sin \theta \, \cos \theta \, \frac{i g }{2 \sqrt{2}(2\pi)^\frac{3}{2}} \, \de^3(\G p-\G q -\G k) \, \int^{x^0_{out}}_{x^0_{in}} \!\! \dr x^0 \frac{\varepsilon_{\G p, \mu, \la}}{\sqrt{2 E^W_p}}\, \de^3(\G p-\G q -\G k) \\[2mm] \non
& \times &  \lf\{ e^{-i \om_{\G k,2} \, x^0_{in}} \, \bar{u}^r_{\G k,2} \, \ga^\mu (1-\ga^5) \, v^s_{\G q,e} \, e^{-i (E^W_p-E_q^e-\om_{\G k,2}) x^0} \ri.	\\[2mm]
& - & 	\lf[|U_\G k| \, e^{-i \om_{\G k,1} \, x^0_{in}} \, \bar{u}^r_{\G k,1} \, \ga^\mu (1-\ga^5) \, v^s_{\G q,e} \, e^{-i (E^W_p-E_q^e-\om_{\G k,1}) x^0}\ri. \non\\[2mm]
 & + & \lf.\lf. \varepsilon^r |V_\G k| \, e^{i \om_{\G k,1} \, x^0_{in}} \, \bar{v}^r_{-\G k,1} \, \ga^\mu (1-\ga^5) \, v^s_{\G q,e} \, e^{-i (E^W_p-E_q^e+\om_{\G k,1}) x^0}\ri]\ri\}\, .
\eea
Looking at the case $\tau_W \ll \De t \ll t_{osc}$, we find~\cite{Blasone:2006jx}:
\bea \non 
\mathcal{A}_{W^+ \rightarrow e^+ \, \nu_e} & \approx & \frac{i g }{2 \sqrt{2}(2\pi)^\frac{3}{2}} \, \de^3(\G p-\G q -\G k)  \frac{\varepsilon_{\G p, \mu, \la}}{\sqrt{2 E^W_p}}\, \de^3(\G p-\G q -\G k) \, \De t\\[2mm]  
& \times &  \lf\{ \cos^2 \theta \, \bar{u}^r_{\G k,2}+\sin^2 \theta 	\lf[|U_\G k| \bar{u}^r_{\G k,2}+ \varepsilon^r |V_\G k| \, \bar{v}^r_{-\G k,2} \ri]\ri\} \, \ga^\mu (1-\ga^5) \, v^s_{\G q,e}\, , 	\label{aee}
\eea
and 
\bea 	\label{aemu}
\mathcal{A}_{W^+ \rightarrow e^+ \, \nu_\mu} & \approx & 0\, .
\eea
which is the expected result.
\section{Conclusions}
We have reviewed some basic facts on neutrino flavor states, and put in evidence some inconsistencies, coming out from the use of weak states which are not eigenstates of the flavor charges. The definition of neutrino flavor states as exact eigenstates of flavor charges \cite{qftmixing} instead provides the solution to overcome such inconsistencies and recovers the correct results, as the exact oscillation formula Eqs. \eqref{fqosc},\eqref{oscfor} and family lepton number conservation (at tree level), in the production/detection vertex Eqs.\eqref{aee},\eqref{aemu}.

Another independent argument in favor of flavor states as ``ontological'' entities, comes from the study of the proton decay in accelerated frames in presence of neutrino mixing \cite{Blasone:2018czm,pos}. From general covariance, the proton decay rates calculated in the inertial and comoving frame should agree: this leads to the necessity of Unruh radiation but also to the use of flavor neutrino states in the decay vertex.

We finally point out that recently \cite{BJSun}, the use of flavor states led to the interpretation of flavor neutrinos as unstable particles, which periodically \emph{decay} into different neutrino species. This view is also compatible with the formal statement, based on the study of Schwinger--Dyson equations, according to which flavor neutrinos can be formally thought as \emph{single-particle bound states}, i.e. bound states in the flavor space.
\section*{Acknowledgments}
L.S. would like to thank G. Fiore, M. Kurkov and F. Lizzi for interesting discussions on the subject.

\section*{References}

\end{document}